\documentclass[twocolumn,pre,superscriptaddress,longbibliography]{revtex4-1}
\usepackage{graphicx,amsmath,amssymb}
\usepackage[colorlinks=true, citecolor=blue, urlcolor=blue, linkcolor=red]{hyperref}

\makeatletter
\begin{document}
\title{Work statistics and thermal phase transitions}

\author{Ze-Zhou Zhang}
\affiliation{Lanzhou Center for Theoretical Physics, Key Laboratory of Theoretical Physics of Gansu Province, Lanzhou University, Lanzhou 730000, China}
\author{Wei Wu}
\email{wuw@lzu.edu.cn}
\affiliation{Lanzhou Center for Theoretical Physics, Key Laboratory of Theoretical Physics of Gansu Province, Lanzhou University, Lanzhou 730000, China}

\begin{abstract}
Many previous studies have demonstrated that work statistics can exhibit certain singular behaviors in the quantum critical regimes of many-body systems at zero or very low temperatures. However, as the temperature increases, it is commonly believed that such singularities will vanish. Contrary to this common recognition, we report a nonanalytic behavior of the averaged work done, which occurs at finite temperature, in the Dicke model as well as the Lipkin-Meshkov-Glick model subjected to the sudden quenches of their work parameters. It is revealed that work statistics can be viewed as a signature of the thermal phase transition when the quenched parameters are tuned across the critical line that separates two different thermal phases.
\end{abstract}
\maketitle

\section{Introduction}\label{sec:sec1}

In recent years, much attention has been focused on the investigation of the nonequilibrium statistical mechanics in quantum systems~\cite{RevModPhys.81.1665,doi:10.1063/1.2012462,RevModPhys.83.771,Seifert_2012,RevModPhys.92.041002,RevModPhys.93.035008,doi:10.1116/5.0083192}. Fluctuation theorems, for example, the Crooks relation~\cite{PhysRevE.60.2721} and the Jarzynski equality~\cite{PhysRevLett.78.2690}, lie at the heart of the nonequilibrium thermodynamics. They establish a bridge connecting the well-defined thermal equilibrium properties, such as the free-energy difference, and certain nonequilibrium probabilities~\cite{PhysRevE.60.2721} or physical quantities~\cite{PhysRevLett.78.2690}. Moreover, these fluctuation theorems are closely related to the time-reversal symmetry, which provides a new insight for us to understand the Second Law of Thermodynamics~\cite{RevModPhys.81.1665,RODUNER20221,PhysRevE.105.024129,PhysRevE.104.L062101,PhysRevE.105.014129,PhysRevLett.109.120604,doi:10.1146/annurev-conmatphys-062910-140506}. Experimental tests of these fluctuation theorems have been reported in Refs.~\cite{doi:10.1126/science.1071152,Collin2005,PhysRevLett.128.040602,PhysRevLett.128.050603,PhysRevLett.109.180601}.

In the context of fluctuation theorems, the notion of work can be defined by calculating the energy difference of a quantum system at the initial and the final times~\cite{PhysRevE.75.050102}. Such a definition is completely different from these of many other articles~\cite{PhysRevE.71.066102,PhysRevLett.93.048302,JARZYNSKI2007495,PhysRevE.78.011116,PhysRevE.88.042136,PhysRevResearch.2.033508}, in which the work is commonly defined as an expectation value of an operator. Using this definition, Talkner \emph{et al}.~\cite{PhysRevE.75.050102} found the work is more like a statistical quantity and all the available statistical information about work is contained in its characteristic function. By far, the work based on the two-point measurement scheme has been widely studied in various systems by using different techniques~\cite{PhysRevLett.124.240603,PhysRevE.101.032111,PhysRevE.101.030101,PhysRevE.100.062119,PhysRevE.94.062133,PhysRevResearch.2.023377,PhysRevResearch.2.023224,PhysRevE.104.034107,PhysRevA.103.042214,G_nay_2022}.

In many previous articles~\cite{PhysRevLett.124.170603,PhysRevLett.101.120603,PhysRevE.88.042109,PhysRevLett.110.135704,PhysRevE.92.032142,PhysRevE.98.022107,Li_2019,PhysRevE.103.032145,PhysRevE.80.061130,PhysRevE.94.052122,PhysRevB.86.161109,PhysRevB.103.144204,PhysRevB.98.104302}, it is found that the averaged work done in a zero temperature sudden quench dynamics of a many-body quantum system displays certain nonanalytic behaviors when the quenched parameters cross over the quantum critical line. Such singularities have been widely reported in several spin-chain systems~\cite{PhysRevLett.101.120603,PhysRevE.88.042109,PhysRevLett.110.135704,PhysRevE.92.032142,PhysRevE.98.022107,Li_2019,PhysRevE.103.032145}, the cavity QED system~\cite{PhysRevE.80.061130,PhysRevE.94.052122}, the Luttinger liquid~\cite{PhysRevB.86.161109}, the conformal field theory model~\cite{PhysRevB.103.144204} and the Haldane model of graphene~\cite{PhysRevB.98.104302}. These results have convincingly demonstrated that the work statistics can be used as a powerful tool to characterize the quantum criticality of a many-body system.

Compared with traditional methods, the work statistics approach to the quantum phase transition has its own superiority, because it requires no prior knowledge about the order parameter or symmetries. However, almost all the existing studies restricted their attentions to the quantum phase transition case, which happens only at zero or very low temperatures. It is commonly believed that, as the temperature increases, the thermal fluctuation generally erases these nonanalytic behaviors of the work statistics~\cite{PhysRevE.92.032142,PhysRevE.98.022107,Li_2019}. This common recognition suggests that the work statistics may lose its ability to characterize the criticality of a many-body system at high temperature.

In this work, we recheck the above common belief by investigating the performance of the averaged work done in a sudden quench nonequilibrium dynamics of a many-body system, which experiences a thermal phase transition purely induced by thermal fluctuations at finite temperature. The Dicke model as well as the Lipkin-Meshkov-Glick model are chosen as the illustrative examples. It is revealed that the averaged work exhibits a singular behavior when the quenched parameters are tuned across the critical boundary that separates two different thermal phases. This result is contrary to the previous common recognition and expands our understanding of the work statistics approach to the criticality of a many-body system. Moreover, the accuracy of our treatments is discussed by evaluating the famous Jarzynski identity.

This paper is organized as follows: In Sec.~\ref{sec:sec2}, we first recall some basic concepts as well as the general formalisms of the work statistics in quantum mechanics. In Sec.~\ref{sec:sec2} (Sec.~\ref{sec:sec3}), we outline the thermodynamic characteristic of the Dicke model (the Lipkin-Meshkov-Glick model) and analyze the performances of the averaged work done in a sudden quench nonequilibrium process of the Dicke model (the Lipkin-Meshkov-Glick model). The effectiveness of our result is analyzed in Sec.~\ref{sec:sec4} by checking the Jarzynski equality. Some discussions and the main conclusions of this paper are drawn in Sec.~\ref{sec:sec5}. In the two appendices, we provide some additional details about the main text. Throughout the paper, we set $k_{\text{B}}=\hbar=1$, and all the other units are dimensionless as well.

\section{Work Statistics}\label{sec:sec2}

In this section, we shall first recall the notion of work in quantum mechanics based on the two-point measurement scheme. Let us consider a quantum system, whose Hamiltonian is described by $\hat{H}(\lambda_{t})$ with $\lambda_{t}$ being a time-dependent externally controllable parameter, evolves from an initially thermal equilibrium state $\rho(0)=e^{-\beta \hat{H}(\lambda_{0})}/\text{Tr}[e^{-\beta \hat{H}(\lambda_{0})}]$ at $t=0$ to a final time $t=\tau$. The parameter $\lambda_{t}$ is also called the work parameter in the nonequilibrium thermodynamics, it drives the quantum system out of equilibrium and injects the energy (work) into the quantum system. The work $W$ performed on the system during the above nonequilibrium process can be quantified by measuring the energy difference of $\hat{H}(\lambda_{t})$ at the initial and the final times. Via decomposing $\hat{H}(\lambda_{t})$ as $\hat{H}(\lambda_{t})=\sum_{i}\varepsilon_{t}^{i}|\varepsilon_{t}^{i}\rangle\langle\varepsilon_{t}^{i}|$, the work distribution function can be expressed as~\cite{PhysRevE.75.050102}
\begin{equation}\label{eq:eq1}
p(W)=\sum_{i,i'}\langle\varepsilon_{0}^{i}|\rho(0)|\varepsilon_{0}^{i}\rangle|\langle\varepsilon_{\tau}^{i'}|\hat{U}(\tau)|\varepsilon_{0}^{i}\rangle|^{2}\delta(W-W_{ii'}),
\end{equation}
where $W_{ii'}\equiv\varepsilon_{\tau}^{i'}-\varepsilon_{0}^{i}$ are the energy differences in two successive measurements, and $\hat{U}(t)\equiv \mathbb{\hat{T}}\exp[-i\int_{0}^{t}d\tau\hat{H}(\lambda_{\tau})]$ denotes the unitary time evolution operator with $\hat{\mathbb{T}}$ being the time ordering operator. Usually, one uses the characteristic function $G(u)$, which is defined as the Fourier transform of $p(W)$~\cite{PhysRevE.75.050102}
\begin{equation}\label{eq:eq2}
\begin{split}
G(u)\equiv&\int dWe^{iuW}p(W)\\
=&\text{Tr}\Big{[}e^{iu\hat{H}(\lambda_\tau)}\hat{U}(\tau)e^{-iu\hat{H}(\lambda_0)}\rho(0)\hat{U}^{\dagger}(\tau)\Big{]},
\end{split}
\end{equation}
to describe the statistical properties of the work. With the expression of $G(u)$ at hand, the averaged work (the first moment) can be calculated as
\begin{equation}\label{eq:eq3}
\langle W\rangle=-i\frac{\partial}{\partial u}G(u)\bigg{|}_{u=0}.
\end{equation}

In this paper, we assume the quantum system undergoes a sudden quench dynamics, which is one of the simplest nonequilibrium processes. In the sudden quench dynamics, the work parameter $\lambda_{t}$ instantaneously changes from the initial value $\lambda_{0}=\lambda_{\text{i}}$ to the final value $\lambda_\tau=\lambda_{\text{f}}$. Such a sudden quench dynamics leads to $\hat{U}(\tau)$ becomes an identity operator, and the characteristic function can be simplified to
\begin{equation}\label{eq:eq4}
G(u)=\text{Tr}\Big{[}e^{iu\hat{H}(\lambda_{\text{f}})}e^{-iu\hat{H}(\lambda_{\text{i}})}\rho(0)\Big{]}.
\end{equation}
The above equation for a sudden quench nonequilibrium process has been widely used in many previous studies~\cite{PhysRevLett.124.170603,PhysRevLett.101.120603,PhysRevE.88.042109,PhysRevLett.110.135704,PhysRevE.92.032142,PhysRevE.98.022107,Li_2019,PhysRevE.103.032145,PhysRevE.80.061130,PhysRevE.94.052122,PhysRevB.86.161109,PhysRevB.103.144204,PhysRevB.98.104302}. Next, by using two well-known quantum many-body models as the illustrative examples, we explore the relation between the work statistics and the thermal phase transition at finite temperature.

\section{The Dicke model case}\label{sec:sec3}

Our first illustrative example is the Dicke model~\cite{PhysRev.93.99}, which describes the interaction between an atomic ensemble with $N$ spins and a single-mode cavity field. The Hamiltonian of the Dicke model is given by
\begin{equation}\label{eq:eq5}
\hat{H}_{\text{DM}}=\epsilon \hat{J}_{z}+\omega \hat{a}^{\dagger}\hat{a}+\frac{2\gamma}{\sqrt{N}}\hat{J}_{x}(\hat{a}^{\dagger}+\hat{a}),
\end{equation}
where $\hat{J}_{z,x}\equiv\frac{1}{2}\sum_{n=1}^{N}\hat{\sigma}_{n}^{z,x}$ are the collective spin operators of the atomic ensemble. The parameter $\epsilon$ denotes the energy splitting induced by an external field. Operators $\hat{a}^{\dagger}$ and $\hat{a}$ are the creation and the annihilation operators of the single-mode cavity field with the corresponding frequency $\omega$, respectively. And the parameter $\gamma$ quantifies the coupling strength between the atomic ensemble and the cavity field. As shown in Refs.~\cite{PhysRevA.70.033808,Liberti2005}, the interaction between the atomic ensemble and the cavity field can be interpreted as an effective spin-spin interaction of a long-range nature. Then, the competition between the above long-range spin-spin interaction and the external field term $\epsilon \hat{J}_{z}$ leads to a second-order thermal phase transition~\cite{PhysRevA.70.033808,Liberti2005,PhysRevA.9.418,PhysRevA.7.831,Bastarrachea_Magnani_2016}.

\subsection{Thermodynamic properties of the Dicke model}

To discuss the thermodynamic properties of the Dicke model, one needs to obtain the expression of the partition function $Z_{\text{DM}}\equiv \text{Tr}(e^{-\beta \hat{H}_{\text{DM}}})$. By employing the same analytical method reported in Refs.~\cite{PhysRevA.9.418,PhysRevA.7.831,Bastarrachea_Magnani_2016}, in the thermodynamic limit $N\rightarrow\infty$, one can find the partition function of the Dicke model can be expressed as (see Appendix A for more details)
\begin{equation}\label{eq:eq6}
Z_{\text{DM}}=\sqrt{\frac{2}{\beta\omega|\partial_{z}^{2}\Phi(z)|}}e^{N \Phi(z)}\Bigg{|}_{z=z_{0}},
\end{equation}
where $\Phi(z)$ is defined by
\begin{equation}\label{eq:eq7}
\Phi(z)\equiv-\beta\omega z^{2}+\ln\Bigg{[}2\cosh\bigg{(}\frac{\beta}{2}\sqrt{\epsilon^{2}+16\gamma^{2}z^{2}}\bigg{)}\Bigg{]},
\end{equation}
and $z_{0}$ is determined by $\phi(z_{0})=0$ with $\phi(z)\equiv\partial_{z}\Phi(z)$. Thus, the thermodynamic properties of the Dicke model is determined by the roots of the equation $\phi(z_{0})=0$. As discussed in Refs.~\cite{PhysRevA.70.033808,Liberti2005,PhysRevA.9.418,PhysRevA.7.831,Bastarrachea_Magnani_2016}, there are two possible roots, depending on the critical temperature
\begin{equation}\label{eq:eq8}
T_{\text{c}}^{\text{DM}}=\epsilon\bigg{[}2\mathrm{arctanh}\bigg{(}\frac{\epsilon\omega}{4\gamma^{2}}\bigg{)}\bigg{]}^{-1}.
\end{equation}
When $T>T_{\text{c}}^{\text{DM}}$, the equation $\phi(z_{0})=0$ has a trivial solution $z_{0}=0$ corresponding especially to the case in which the atomic ensemble and the cavity field are completely decoupled. On the other hand, if $T\leq T_{\text{c}}^{\text{DM}}$, a nontrivial solution $z_{0}=\sqrt{\epsilon^{2}\eta^{2}-\epsilon^{2}}/(4\gamma)$ with $\eta$ determined by $\frac{1}{4}\eta\epsilon\omega\gamma^{-2}=\tanh(\frac{1}{2}\beta\eta\epsilon)$ can be found. From the above analysis, one can conclude that the Dicke model experiences a thermodynamic phase transition at the critical temperature $T=T_{\text{c}}^{\text{DM}}$. Above the critical temperature $T_{\text{c}}^{\text{DM}}$, the Dicke model is in the normal phase (NP). However, if $T\leq T_{\text{c}}^{\text{DM}}$, the Dicke model is in the superradiant phase (SP). Moreover, in the limit $T_{\text{c}}^{\text{DM}}\rightarrow 0$, one can find Eq.~(\ref{eq:eq8}) reduces to $\gamma_{\text{c}}=\frac{1}{2}\sqrt{\epsilon\omega}$ which is in agreement with the critical coupling strength of emerging the quantum phase transition in the Dicke model at zero temperature~\cite{PhysRevE.67.066203}.

In Figs.~\ref{fig:fig1}-(a) and (b), we plot the thermal phase diagram obtained by Eq.~(\ref{eq:eq8}) as well as the specific heat capacity per atom $C/N$ of the Dicke model, respectively. The explicit expression of $C/N$ is given in Appendix A. From Fig.~\ref{fig:fig1}-(b), one can see $C/N$ exhibits two completely different thermodynamic behaviors in the normal and superradiant phases. This result means the first derivative of $C/N$ exhibits a singular behavior when it crosses the thermal phase boundary and verifies our previous analysis based on the partition function.

\subsection{Work statistics and thermal phase transition of the Dicke model}

\begin{figure}
\centering
\includegraphics[angle=0,width=8.85cm]{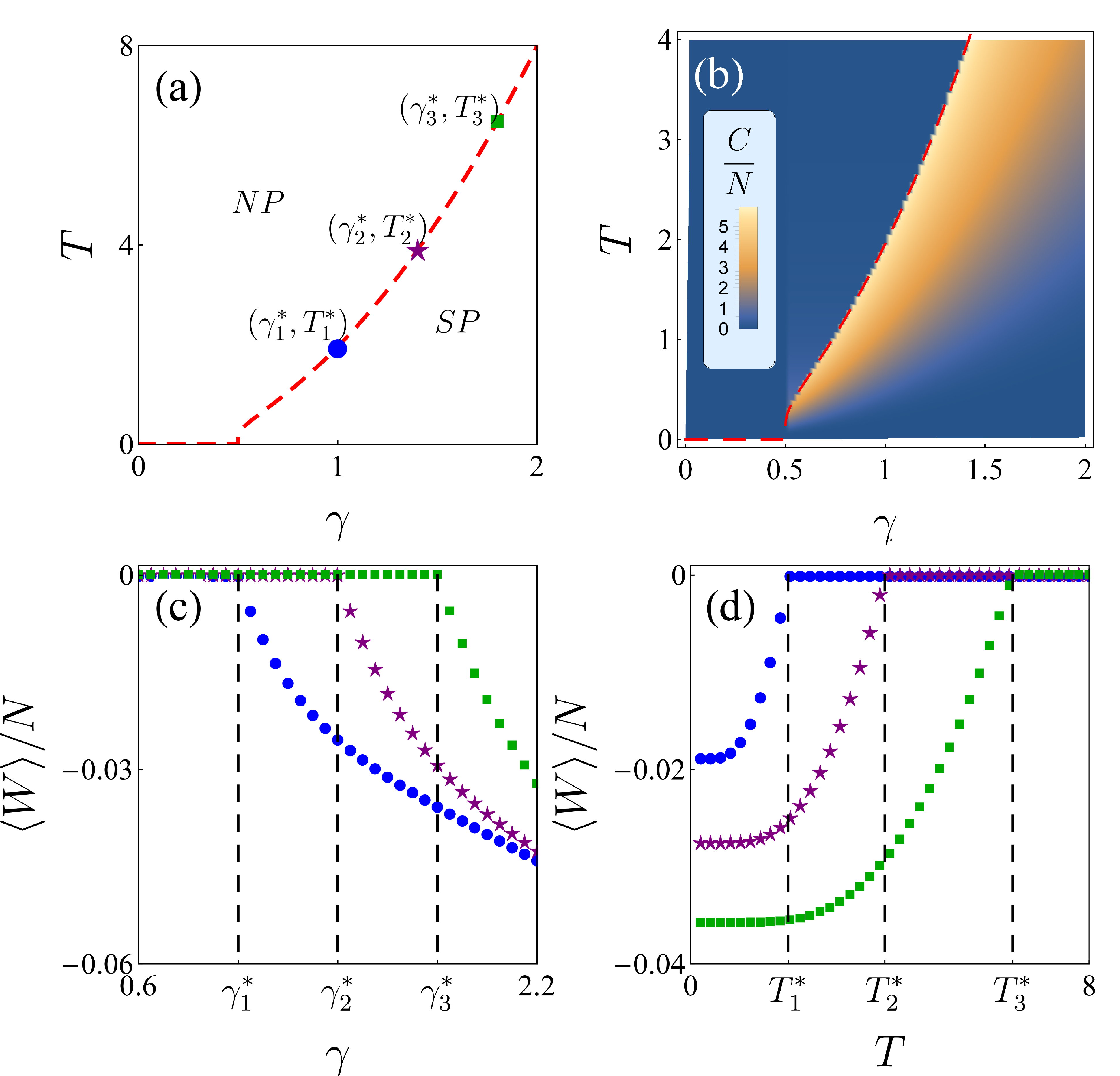}
\caption{(a) The thermodynamic phase diagram of the Dicke model obtained by Eq.~(\ref{eq:eq8}). (b) The specific heat capacity per atom $C/N$ versus the temperature $T$ and the coupling strength $\gamma$. The red dashed lines in (a) and (b) mark the boundary between the NP regime and the SP regime. Here, three critical points $(\gamma_{\nu}^{*},T_{\nu}^{*})$ with $\nu=1,2,3$ along the phase boundary are chosen as the examples to show the relation between the averaged work and the thermal phase transition of the Dicke model. (c) The averaged work $\langle W\rangle$ is plotted as a function of the coupling strength $\gamma$ with fixed temperatures. One can see the averaged work becomes discontinuous when it crosses over the critical coupling $\gamma_{\nu}^{*}$ with fixed temperature $T_{\nu}^{*}$. (d) The averaged work $\langle W\rangle$ is plotted as a function of $T$ with fixed the coupling strengths. One can see $\langle W\rangle$ becomes discontinuous when it crosses over the critical temperature $T_{\nu}^{*}$ with respect to the given $\gamma_{\nu}^{*}$. The parameters are chosen as $\epsilon=1$, $\omega=1$, $\delta=0.01$, $\gamma_{1}^{*}=1$ (blue circle), $\gamma_{2}^{*}=1.4$ (purple star) and $\gamma_{3}^{*}=1.8$ (green rectangle).}\label{fig:fig1}
\end{figure}

In the Dicke model case, the coupling strength $\gamma$ is chosen as the work parameter, which rapidly changes from $\lambda_{\text{i}}=\gamma$ to $\lambda_{\text{f}}=\gamma+\delta$. In the limit $\delta/\gamma\rightarrow0$, the characteristic function given by Eq.~(\ref{eq:eq4}) can be approximately derived as follows
\begin{equation}\label{eq:eq9}
\begin{split}
G(u)=&\frac{1}{Z_{\text{DM}}}\text{Tr}\Big{[}e^{iu(\hat{H}_{\text{DM}}+\delta \hat{v})}e^{-(\beta+iu)\hat{H}_{\text{DM}}}\Big{]}\\
\simeq&\frac{1}{Z_{\text{DM}}}\text{Tr}\Big{[}e^{iu\hat{H}_{\text{DM}}}e^{iu\delta \hat{v}}e^{-(\beta+iu)\hat{H}_{\text{DM}}}\Big{]}\\
=&\frac{1}{Z_{\text{DM}}}\text{Tr}\Big{(}e^{iu\delta \hat{v}}e^{-\beta\hat{H}_{\text{DM}}}\Big{)}\\
=&\overline{e^{iu\delta \hat{v}}}\simeq e^{iu\delta\overline{\hat{v}}-\frac{1}{2}u^{2}\delta^{2}(\overline{\hat{v}^{2}}-\overline{\hat{v}}^{2})},
\end{split}
\end{equation}
where $\hat{v}\equiv\frac{2}{\sqrt{N}}\hat{J}_{x}(\hat{a}^{\dagger}+\hat{a})$ and $\overline{\hat{o}}\equiv Z_{\text{DM}}^{-1}\text{Tr}(\hat{o}e^{-\beta\hat{H}_{\text{DM}}})$ denotes the thermodynamic averaged value with respect to the thermal Gibbs state of the Dicke model. The explicit expressions of $\overline{\hat{v}}$ and $\overline{\hat{v}^{2}}$ are given in Appendix B. With Eq.~(\ref{eq:eq9}) at hand, we find the expression of averaged work done in the above sudden quench process is given by
\begin{equation}\label{eq:eq10}
\langle W\rangle=-\frac{4N\gamma\delta z_{0}^{2}}{\sqrt{\epsilon^{2}+16\gamma^{2}z_{0}^{2}}}\tanh\bigg{(}\frac{\beta}{2}\sqrt{\epsilon^{2}+16\gamma^{2}z_{0}^{2}}\bigg{)}.
\end{equation}
From the above expression, one can immediately find $\langle W\rangle=0$ in the NP with $z_{0}=0$ and $\langle W\rangle< 0$ in the SP with $z_{0}>0$. This result means the averaged work can be regarded as an order parameter to reveal the thermal phase transition of the Dicke model. In Figs.~\ref{fig:fig1}(c) and (d), the averaged work $\langle W\rangle$ is plotted as a function of the coupling strength $\gamma$ and the temperature $T$, respectively. One can see $\langle W\rangle$ exhibits a discontinuous behavior when crossing over the thermal phase boundary. Such a singularity is quite similar to previous studies at zero temperature~\cite{PhysRevLett.101.120603,PhysRevE.88.042109,PhysRevLett.110.135704,PhysRevE.92.032142,PhysRevE.98.022107,Li_2019,PhysRevE.103.032145,PhysRevE.80.061130,PhysRevE.94.052122} and can be used to reveal the thermal phase transition without a prior knowledge about the order parameter or symmetries.

\section{The Lipkin-Meshkov-Glick model case}\label{sec:sec4}

Our second example is the Lipkin-Meshkov-Glick model~\cite{LIPKIN1965188}, which describes a collective spin in an external magnetic field. The Hamiltonian of the Lipkin-Meshkov-Glick model is described by
\begin{equation}\label{eq:eq11}
\hat{H}_{\text{LMG}}=-\chi\hat{J}_{z}-\frac{1}{N}\hat{J}_{x}^{2},
\end{equation}
where $\chi$ is the strength of the applied external field. Similar to that of the Dicke model case, the competition between the spin-spin interaction $\hat{J}_{x}^{2}$ and the effect of the external field $-\chi\hat{J}_{z}$ gives rise to a second-order thermal phase transition from the paramagnetic phase (PP) to the ferromagnetic phase (FP)~\cite{LIPKIN1965188,PhysRevE.79.031101,Wilms_2012}.

\subsection{Thermodynamic properties of the Lipkin-Meshkov-Glick model}

\begin{figure}
\centering
\includegraphics[angle=0,width=8.75cm]{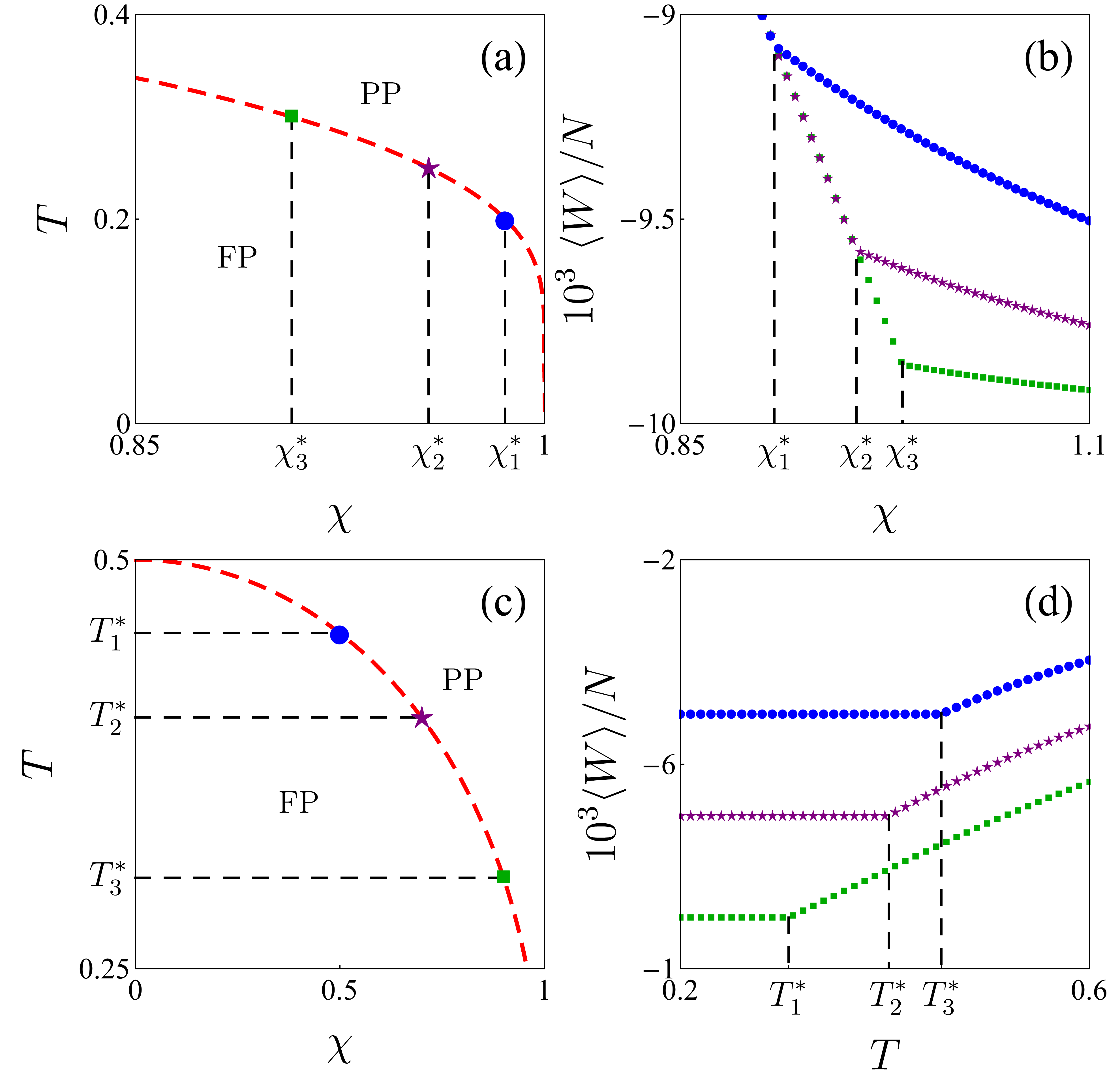}
\caption{(a) The thermal phase diagram of the Lipkin-Meshkov-Glick model. Along the phase boundary, three critical points with $\chi_{\nu}^{*}$ are chosen as the examples, which are marked by blue circles, purple stars and green rectangles, respectively. (b) The averaged work $\langle W\rangle$ is plotted as a function of $\chi$ with fixed temperatures: $T_{1}=0.2$ (blue circles), $T_{2}=0.25$ (purple stars) and $T_{3}=0.3$ (green rectangles). One can see the averaged work becomes discontinuous when it crosses over $\chi_{\nu}^{*}$ with the fixed temperature $T_{\nu}$. (c) The same with (b), but the critical parameters are chosen as $T_{\nu}^{*}$. (d) The averaged work $\langle W\rangle$ is plotted as a function of $T$ with fixed coupling strengths: $\chi_{1}=0.5$ (blue circles), $\chi_{2}=0.7$ (purple stars), and $\chi_{3}=0.9$ (green rectangles). A singular behavior is also found if $T$ is tuned across $T_{\nu}^{*}$ with respect to the fixed $\chi_{\nu}$. The parameter of $\delta$ is chosen as $\delta=0.02$.}\label{fig:fig2}
\end{figure}

The Lipkin-Meshkov-Glick Hamiltonian can be exactly treated by making use of certain numerical techniques~\cite{PhysRevB.74.104118,PhysRevLett.99.050402}. However, to obtain an analytical result with a clear physical picture, in this paper, we apply the standard mean-field approximation~\cite{PhysRevE.79.031101,Wilms_2012} to the Lipkin-Meshkov-Glick model. To this aim, we shall first introduce a quantity
\begin{equation}\label{eq:eq12}
\Omega\equiv\frac{2}{N}\text{Tr}\bigg{(}\hat{J}_{x}\frac{e^{-\beta \hat{H}_{\text{LMG}}}}{Z_{\text{LMG}}}\bigg{)},
\end{equation}
which is the average magnetization along the $x$ direction, and reexpress each individual Pauli-$x$ spin operator $\hat{\sigma}_{n}^{x}$ as $\hat{\sigma}_{n}^{x}=\Omega+(\hat{\sigma}_{n}^{x}-\Omega)$. Then, by plugging this expression into the original Hamiltonian of the Lipkin-Meshkov-Glick model, we have
\begin{equation}\label{eq:eq13}
\hat{H}_{\text{LMG}}=-\frac{1}{4}\bigg{[}2\Omega\sum_{n}\hat{\sigma}_{n}^{x}+\sum_{n}(\hat{\sigma}_{n}^{x}-\Omega)^{2}-\Omega^{2}\bigg{]}-\chi \hat{J}_{z}.
\end{equation}
Neglecting the fluctuations involving $(\hat{\sigma}_{n}^{x}-\Omega)^{2}$ as well as all the higher order terms $\mathcal{O}(N^{-2})$, an effective Hamiltonian under the mean-field approximation can be derived~\cite{PhysRevE.79.031101,Wilms_2012}:
\begin{equation}\label{eq:eq14}
\hat{H}_{\text{LMG}}^{\text{MF}}=-\Omega_{\text{MF}}\hat{J}_{x}-\chi \hat{J}_{z},
\end{equation}
where $\Omega_{\text{MF}}$ is the value of $\Omega$ within the mean-field treatment and will be determined later. One can find that the above mean-field Hamiltonian is a sum of decoupled single-spin Hamiltonians, thus it can be diagonalized directly.

The partition function under the mean-field approximation can be easily obtained as
\begin{equation}\label{eq:eq15}
\begin{split}
Z_{\text{LMG}}^{\text{MF}}=&\text{Tr}\Big{(}e^{-\beta\hat{H}_{\text{LMG}}^{\text{MF}}}\Big{)}=\Big{[}2\cosh(\beta\Theta_{0})\Big{]}^{N},
\end{split}
\end{equation}
where $\Theta_{x}\equiv\frac{1}{2}\sqrt{(\chi+x)^{2}+\Omega_{\text{MF}}^{2}}$. Using Eq.~(\ref{eq:eq12}) and Eq.~(\ref{eq:eq15}), one can find the following self-consistent equation of $\Omega_{\text{MF}}$:
\begin{equation}\label{eq:eq16}
\Omega_{\text{MF}}=\frac{\tanh(\beta\Theta_{0})}{2\Theta_{0}}\Omega_{\text{MF}}.
\end{equation}
By analyzing the roots of the above equation, the thermal phase diagram of the Lipkin-Meshkov-Glick model can be obtained. When $T>T_{\text{c}}^{\text{LMG}}$, one possible solution of Eq.~(\ref{eq:eq16}) is $\Omega_{\text{MF}}=0$, which implies the Lipkin-Meshkov-Glick model is in the PP regime. On the contrary, if $T\leq T_{\text{c}}^{\text{LMG}}$, a non-zero averaged magnetization along the $x$ direction can be found, which satisfies $2\Theta_{0}=\tanh[\Theta_{0}/(2T)]$ and means the Lipkin-Meshkov-Glick model is in the FP regime. The critical temperature $T_{\text{c}}^{\text{LMG}}$ is given by~\cite{PhysRevE.79.031101,Wilms_2012}
\begin{equation}\label{eq:eq17}
T_{\text{c}}^{\text{LMG}}=\frac{\chi}{2\text{arctanh}(\chi)}.
\end{equation}
In the limit $T_{\text{c}}^{\text{LMG}}\rightarrow0$, the quantum phase transition point of the Lipkin-Meshkov-Glick model $\chi_{\text{c}}=1$~\cite{PhysRevLett.99.050402} can be naturally recovered. The thermal phase diagram based on the above analysis is displayed in Figs.~\ref{fig:fig2}-(a) and (c).

\subsection{Work statistics and thermal phase transition of the Lipkin-Meshkov-Glick model}

Similar to that of the Dicke model case, we chose the external field strength as the work parameter which changes from $\lambda_{\text{i}}=\chi$ to $\lambda_{\text{f}}=\chi+\delta$ in the sudden quench process. With the mean-field Hamiltonian $\hat{H}_{\text{LMG}}^{\text{MF}}$ at hand, we find the characteristic function is given by
\begin{equation}\label{eq:eq18}
\begin{split}
G(u)\simeq&\frac{1}{Z^{\text{MF}}_{\text{LMG}}}\text{Tr}\Big{[}e^{iu(\hat{H}^{\text{MF}}_{\text{LMG}}+\delta \hat{J}_{z})}e^{-(\beta+iu)\hat{H}^{\text{MF}}_{\text{LMG}}}\Big{]}\\
=&\frac{1}{Z_{\text{LMG}}^{\text{MF}}}  \bigg{\{}2\cos(u\Theta_{\delta}){\rm cos}[(u-i\beta)\Theta_{0}]\\
&+2{\rm sin}(u\Theta_{\delta}){\rm sin}[(u-i\beta)\Theta_{0}] \frac{4\Theta_{0}^{2}+\chi\delta}{4\Theta_{0}\Theta_{\delta}} \bigg{\}}^{N},
\end{split}
\end{equation}
which results in
\begin{equation}\label{eq:eq19}
\langle W\rangle=-\frac{N\chi\delta}{4\Theta_{0}}\tanh\bigg{(}\frac{\Theta_{0}}{T}\bigg{)},
\end{equation}
for the Lipkin-Meshkov-Glick model case. In Figs.~\ref{fig:fig2} (b) and (d), we plot the averaged work $\langle W\rangle/N$ as functions of $\chi$ and $T$, respectively. A discontinuous behavior is clearly found when the quenched parameter is tuned across the thermal critical line. This result demonstrates the work statistics can be also viewed as a good detector of the thermal phase transition occurring in the Lipkin-Meshkov-Glick model as well.

\section{Evaluation of the Jarzynski equality}\label{sec:sec5}

\begin{figure}
\centering
\includegraphics[angle=0,width=5.5cm]{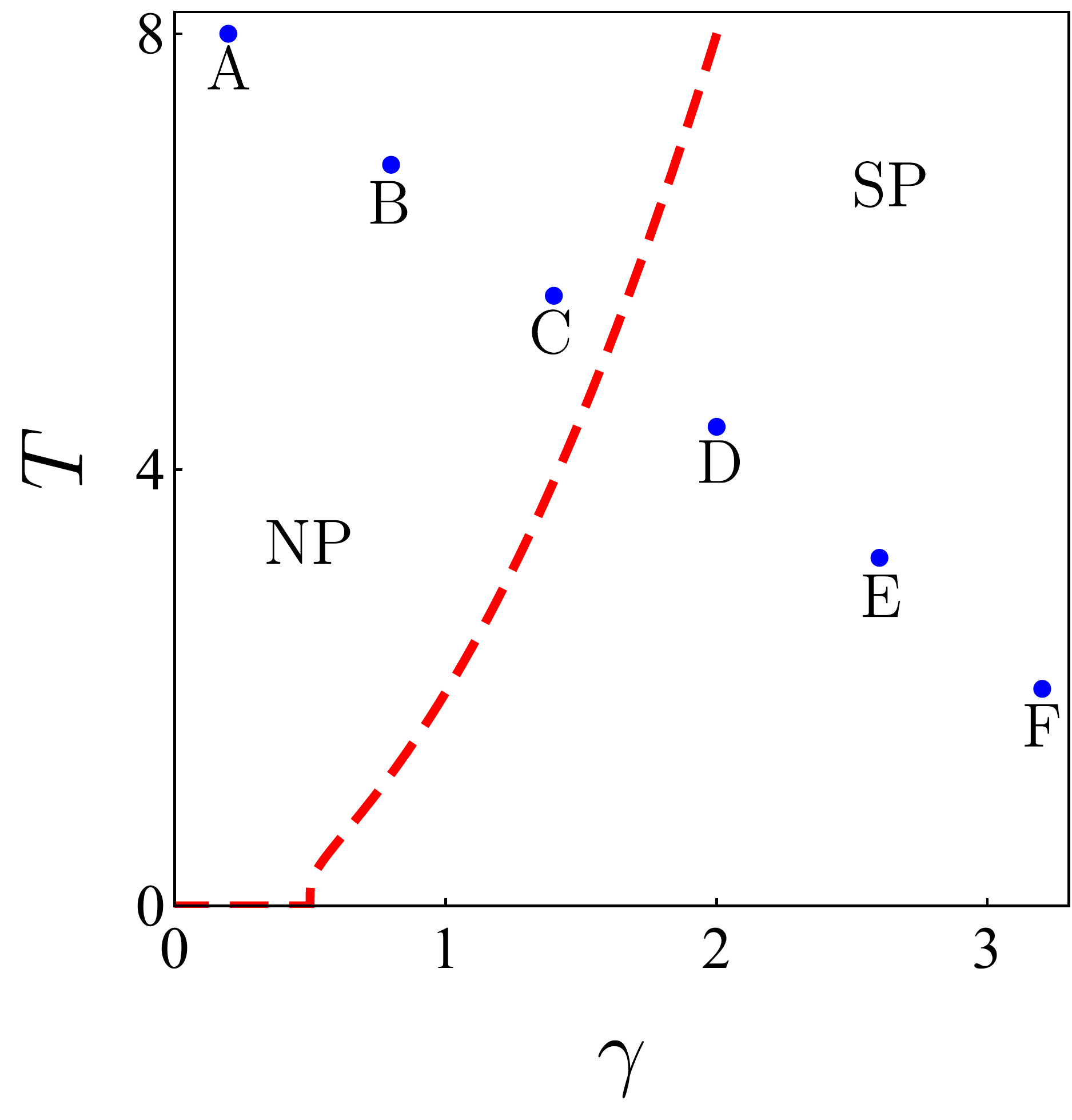}
\caption{The thermal phase diagram of the quantum Dicke model with six representative data in the parameter space $(\gamma,T)$: A (0.2,8), B (0.8,6.8), C (1.4,5.6), D (2,4.4), E (2.6,3.2) and F (3.2,2). These parameters are used to evaluate the Jarzynski identity in Table~\ref{table:table1}. Other parameters are $\epsilon=\omega=1$.}\label{fig:fig3}
\end{figure}

\begin{table}[t]
\begin{center}
\caption{The Jarzynski identity is checked with 6 representative parameters A-F marked by blue circles in Fig.~\ref{fig:fig3}. Parameters are $\delta=10^{-4}$ and $N=10^{2}$.}\label{table:table1}
\setlength{\tabcolsep}{4.75pt}
\begin{tabular}{ccccccc}
  \hline
  \hline
  \text{Parameters} & A & B & C & D & E & F \\
  \hline
  $\langle e^{-\beta W+\beta\Delta F}\rangle$ & 1.000 & 1.000 & 1.000 & 0.992 & 0.984 & 0.969 \\
  \hline
  \hline
\end{tabular}
\end{center}
\end{table}

To obtain Eq.~(\ref{eq:eq9}), we have used the assumptions of $\delta/\gamma\rightarrow 0$ and $N\rightarrow\infty$. On the other hand, to derive Eq.~(\ref{eq:eq18}), we have employed the mean-field approximation which is acceptable only in the case $N\rightarrow\infty$. A natural question is whether or not these approximate treatments are reliable. To address the above question, in this section, we benchmark the accuracy of our results by evaluating the famous Jarzynski identity, which is independent of the protocol generating the nonequilibrium dynamics.

The Jarzynski equality states that~\cite{PhysRevLett.78.2690}
\begin{equation}\label{eq:eq20}
\langle e^{-\beta W}\rangle=e^{-\beta\Delta F},
\end{equation}
where $\langle e^{-\beta W}\rangle\equiv\int dWe^{-\beta W}p(W)$ denotes an ensemble average of the nonequilibrium exponential work, and $\Delta F$ is the free energy difference of the quantum system between the initial and the final times. One can easily check that $\langle e^{-\beta W}\rangle=G(i\beta)$ and $e^{-\beta\Delta F}=Z_{\text{f}}/Z_{\text{i}}$ for the sudden quench nonequilibrium process considered in this paper.

Our analytical expressions for the characteristic function and the partition function of the Dicke model have been already given by Eq.~(\ref{eq:eq9}) and Eq.~(\ref{eq:eq6}), respectively. Using these equations, the correctness of our approach can be checked by evaluating the Jarzynski equality. As shown in Fig.~\ref{fig:fig3} and Table~\ref{table:table1} with six representative parameters, we find, in the region where the approximations are valid $(\delta/\gamma\rightarrow 0$ and $N\rightarrow\infty)$, the Jarzynski identity in the Dicke model case is substantially satisfied with a reasonable error.

For the Lipkin-Meshkov-Glick model case, the verification of the Jarzynski equality is more straightforward. From Eq.~(\ref{eq:eq18}), one can immediately find
\begin{equation}\label{eq:eqadd}
\begin{split}
G(i\beta)=&\frac{[2\cos(i\beta\Theta_{\delta})]^{N}}{Z_{\text{LMG}}^{\text{MF}}}\\
=&\frac{[2\cosh(\beta\Theta_{\delta})]^{N}}{[2\cosh(\beta\Theta_{0})]^{N}}=\frac{Z_{\text{f}}}{Z_{\text{i}}}.
\end{split}
\end{equation}
The above equation means the Jarzynski equality can be exactly satisfied in the Lipkin-Meshkov-Glick model case. The results from Table~\ref{table:table1} and Eq.~(\ref{eq:eqadd}) convince us that our treatment is physically acceptable in spite of certain approximations being employed.

\section{Discussion and Conclusion}\label{sec:sec6}

It is necessary to emphasize that our present results are utterly different from some previous studies of the work statistics in the sudden quench dynamics of spin-chains at finite temperature~\cite{PhysRevE.92.032142,PhysRevE.98.022107,Li_2019}. Although the effect of temperature has been taken into account, these studies essentially concentrated on the relation between the work statistics and the quantum criticality in very low temperature regions. The singular behaviors of the work statistics at the quantum phase point are intrinsically induced by the quantum fluctuation. With the increase of temperature, the quantum fluctuation becomes weak and is ultimately washed out at high temperature~\cite{PhysRevE.92.032142,PhysRevE.98.022107,Li_2019}. In sharp contrast to these previous references, in our paper, the singular behaviors of the work statistics purely root from the thermal fluctuations near the phase transition point, which shall not vanish at high temperature. In this sense, our results greatly enrich the scope of the work statistics approach to the criticality of a quantum many-body system.

In summary, we investigate the statistics of the work done in a sudden quench nonequilibrium dynamics of the Dicke model and the Lipkin-Meshkov-Glick model, which display thermal phase transitions at finite temperature. It is revealed that the averaged work exhibits a singular behavior when the quenched parameters are tuned across the critical boundary that separates two different thermal phases. This result is verified by evaluating the Jarzynski identity and means the work statistics can be employed to characterize thermal phase transitions of quantum many-body systems. We expect our results to be of interest for the nonequilibrium statistical mechanics in quantum many-body systems.

\section{Acknowledgments}\label{sec:sec7}

The authors thank Prof. Jun-Hong An and Prof. Hong-Gang Luo for many fruitful discussions. This work was supported by the National Natural Science Foundation (Grants No. 11704025 and No. 12047501).

\begin{widetext}

\section{Appendix A: The partition function of the Dicke model}\label{sec:secappa}

The Hamiltonian of Eq.~(\ref{eq:eq5}) can be rewritten as
\begin{equation}\label{eq:eq21}
\hat{H}_{\text{DM}}=\sum_{n=1}^{N}\hat{H}_{\text{DM}}^{n}=\sum_{n=1}^{N}\bigg{[}\omega \frac{\hat{a}^{\dagger}}{\sqrt{N}}\frac{\hat{a}}{\sqrt{N}}
    +\frac{\epsilon}{2}\hat{\sigma}_{z}^{n}+\frac{\gamma}{\sqrt{N}}(\hat{a}+\hat{a}^{\dagger})\hat{\sigma}_{x}^{n}\bigg{]}.
\end{equation}
Then, the partition function can be calculated as
\begin{equation}\label{eq:eq22}
\begin{split}
Z_{\text{DM}}=\text{Tr}\Big{(}e^{-\beta\hat{H}_{\text{DM}}}\Big{)}=\sum_{\sigma_{1}=\uparrow\downarrow}\sum_{\sigma_{2}=\uparrow\downarrow}...\sum_{\sigma_{N}=\uparrow\downarrow}\langle\sigma_{1}\sigma_{2}...\sigma_{N}|\int_{-\infty}^{+\infty} \frac{d^{2}\alpha}{\pi}\langle\alpha|e^{-\beta\hat{H}_{\text{DM}}}|\alpha\rangle|\sigma_{1}\sigma_{2}...\sigma_{N}\rangle,
\end{split}
\end{equation}
where $|\alpha\rangle$ is the coherent state, $|\uparrow\rangle$ and $|\downarrow\rangle$ are spin-up and spin-down states, respectively. In the limit $N\rightarrow\infty$, we have $\sqrt{N}\gg\max\{\omega,\gamma\}$ which leads to
\begin{equation}\label{eq:eq23}
\begin{split}
\langle\alpha|e^{-\beta\hat{H}_{\text{DM}}}|\alpha\rangle\simeq\prod_{n}\langle\alpha|e^{-\beta \hat{H}^{n}_{\text{DM}}}|\alpha\rangle\simeq\prod_{n}e^{-\beta \langle\alpha|\hat{H}^{n}_{\text{DM}}|\alpha\rangle}=e^{-\beta|\alpha|^{2}}\prod_{n}e^{-\beta \hat{\mathcal{H}}^{n}_{\text{DM}}},
\end{split}
\end{equation}
where
\begin{equation}\label{eq:eq24}
\hat{\mathcal{H}}^{n}_{\text{DM}}=\frac{\epsilon}{2}\hat{\sigma}_{z}^{n}+\frac{2\gamma\text{Re}\alpha}{\sqrt{N}}\hat{\sigma}_{x}^{n}.
\end{equation}
Thus, we have
\begin{equation}\label{eq:eq25}
\begin{split}
Z_{\text{DM}}\simeq&\int_{-\infty}^{+\infty} \frac{d^{2}\alpha}{\pi}e^{-\beta|\alpha|^{2}}\bigg{(}\sum_{\sigma=\uparrow\downarrow}\langle\sigma|e^{-\beta\hat{\mathcal{H}}^{n}_{\text{DM}}}|\sigma\rangle\bigg{)}^{N}\\
=&\int_{-\infty}^{+\infty} \frac{d^{2}\alpha}{\pi}e^{-\beta|\alpha|^{2}}\Bigg{\{}2\cosh\Bigg{[}\beta\sqrt{\frac{\epsilon^{2}}{4}+\frac{4\gamma^{2}(\mathrm{Re}\alpha)^{2}}{N}}\Bigg{]}\Bigg{\}}^{N}.
\end{split}
\end{equation}
To handle the $d^{2}\alpha$-integral, we introduce $x\equiv\mathrm{Re}\alpha$ and $y\equiv\mathrm{Im}\alpha$, which means $d^{2}\alpha=dxdy$ and $|\alpha|^{2}=x^{2}+y^{2}$. By doing so, the $y$-part of the integral can be immediately carried out, and then one can find
\begin{equation}\label{eq:eq26}
\begin{split}
Z_{\mathrm{DM}}=&\frac{1}{\sqrt{\pi\beta\omega}}\int_{-\infty}^{\infty}dxe^{-\beta\omega x^{2}}\Bigg{[}2\cosh\Bigg{(}\beta\sqrt{\frac{\epsilon^{2}}{4}+\frac{4\gamma^{2}x^{2}}{N}}\Bigg{)}\Bigg{]}^{N}.
\end{split}
\end{equation}
The above expression is still intricate. We use the steepest descent method or the so-called Laplace's integral method~\cite{PhysRevA.70.033808,Liberti2005,PhysRevA.9.418,PhysRevA.7.831,Bastarrachea_Magnani_2016} to further simplify the above expression. To this aim, we replace $x/\sqrt{N}$ by a new variable $z$, then the expression of $Z_{\mathrm{DM}}$ can be rewritten as
\begin{equation}\label{eq:eq27}
\begin{split}
Z_{\mathrm{DM}}=&\sqrt{\frac{N}{\pi\beta\omega}}\int_{-\infty}^{\infty}dze^{N\Phi(z)},
\end{split}
\end{equation}
where $\Phi(z)$ is given by Eq.~(\ref{eq:eq7}) in the main text. The form of the partition function in Eq.~(\ref{eq:eq27}) is especially suitable for the Laplace's integral method, which consists in approximating the exponential integrand by a Gaussian function around the global maximum of the function $\Phi(z)$. By employing the Laplace approximation, one can finally obtain~\cite{PhysRevA.70.033808,Liberti2005,PhysRevA.9.418,PhysRevA.7.831,Bastarrachea_Magnani_2016}
\begin{equation}\label{eq:eq28}
\begin{split}
Z_{\mathrm{DM}}\simeq\sqrt{\frac{2}{\beta\omega|\partial_{z}^{2}\Phi(z)|}}e^{N\Phi(z)}\bigg{|}_{z=z_{0}},
\end{split}
\end{equation}
where $z_{0}$ is determined by $\phi(z_{0})=0$. With the expression of $Z_{\mathrm{DM}}$ at hand, one can easily find specific heat per atom is given by
\begin{equation}\label{eq:eq29}
\frac{C}{N}=\frac{\beta^{2}}{N}\frac{\partial^{2}}{\partial \beta^{2}}\ln Z_{\text{DM}}=\bigg{(}\frac{\beta\epsilon}{2}\bigg{)}^{2}\text{sech}\bigg{[}\frac{\beta\xi(z_{0})}{2}\bigg{]}\bigg{\{}1+\frac{16\gamma^{4}}{\epsilon^{2}\omega^{2}}\frac{\text{tanh}[\frac{1}{2}\beta\xi(z_{0})]^{2}}{1-\frac{2\beta\gamma^{2}}{\omega}\text{sech}[\frac{1}{2}\beta\xi(z_{0})]}\delta_{z_{0},0}\bigg{\}},
\end{equation}
where $\xi(x)\equiv\sqrt{\epsilon^{2}+16\gamma^{2}x^{2}}$.

\section{Appendix B: The expressions of $\overline{\hat{v}}$ and $\overline{\hat{v}^{2}}$}\label{sec:secappb}

To find the characteristic function with respect to the Dicke model, one needs the expressions of $\overline{\hat{v}}$ and $\overline{\hat{v}^{2}}$. Using the same method displayed in Appendix A, one can find
\begin{equation}\label{eq:eq30}
\begin{split}
\overline{\hat{v}}=&\frac{1}{Z_{\text{DM}}}\frac{2}{\sqrt{N}}\text{Tr}\Big{[}\hat{J}_{x}(\hat{a}^{\dagger}+\hat{a})e^{-\beta \hat{H}_{\text{DM}}}\Big{]}\\
\simeq&\frac{2\sqrt{N}}{Z_{\text{DM}}}\int_{-\infty}^{+\infty}\frac{d^{2}\alpha}{\pi}e^{-|\alpha|^{2}}\text{Re}\alpha\sum_{\sigma=\uparrow\downarrow}\langle\sigma|\hat{\sigma}_{x}e^{-\beta\hat{\mathcal{H}}_{\text{DM}}}|\sigma\rangle \bigg{[}2\cosh\bigg{(}\frac{\beta\xi_{\alpha}}{2}\bigg{)}\bigg{]}^{N-1}\\
=&-\frac{8\gamma}{Z_{\text{DM}}}\int_{-\infty}^{+\infty}\frac{d^{2}\alpha}{\pi}\frac{(\text{Re}\alpha)^{2}e^{-|\alpha|^{2}}}{\xi_{\alpha}}\sinh\bigg{(}\frac{\beta\xi_{\alpha}}{2}\bigg{)}\bigg{[}2\cosh\bigg{(}\frac{\beta\xi_{\alpha}}{2}\bigg{)}\bigg{]}^{N-1}\\
=&-\frac{4\gamma}{Z_{\text{DM}}}\int_{-\infty}^{+\infty}\frac{d^{2}\alpha}{\pi}\exp\bigg{\{}-|\alpha|^{2}+N\ln\bigg{[}2\cosh\bigg{(}\frac{\beta\xi_{\alpha}}{2}\bigg{)}\bigg{]}\bigg{\}} \frac{(\text{Re}\alpha)^{2}}{\xi_{\alpha}}\tanh\bigg{(}\frac{\beta\xi_{\alpha}}{2}\bigg{)},
\end{split}
\end{equation}
where $\xi_{\alpha}\equiv\xi(\text{Re}\alpha/\sqrt{N})$. The above expression has the same structure with that of Eq.~(\ref{eq:eq25}), which means it can be simplified by applying the Laplace approximation. Using the same process of deriving the partition function displayed in Appendix A, we find
\begin{equation}\label{eq:eq31}
\overline{\hat{v}}\simeq-\frac{4N\gamma z_{0}^{2}}{\xi(z_{0})}\text{tanh}\bigg{[}\frac{1}{2}\beta\xi(z_{0})\bigg{]}.
\end{equation}
and
\begin{equation}\label{eq:eq32}
\overline{\hat{v}^{2}}=16N(N-1)\bigg{\{}\frac{\gamma z_{0}^{2}}{\xi(z_{0})}\tanh\bigg{[}\frac{1}{2}\beta\xi(z_{0})\bigg{]}\bigg{\}}^{2}+Nz_{0}^{2}.
\end{equation}

\end{widetext}

\bibliography{reference}

\end{document}